# Option pricing in a dynamic Variance-Gamma model

Lorenzo Mercuri[1]   Fabio Bellini[2]

We present a discrete time stochastic volatility model in which the conditional distribution of the logreturns is a Variance-Gamma, that is a normal variance-mean mixture with Gamma mixing density.

We assume that the Gamma mixing density is time varying and follows an affine Garch model, trying to capture persistence of volatility shocks and also higher order conditional dynamics in a parsimonious way.

We select an equivalent martingale measure by means of the conditional Esscher transform as in Buhlmann et al. (1996) and show that this change of measure leads to a similar dynamics of the mixing distribution.

The model admits a recursive procedure for the computation of the characteristic function of the terminal logprice, thus allowing semianalytical pricing as in Heston and Nandi (2000).

From an empirical point of view, we check the ability of this model to calibrate SPX option data and we compare it with the Heston and Nandi (2000) model and with the Christoffersen, Heston and Jacobs (2006) model, that is based on Inverse Gaussian innovations. Moreover, we provide a detailed comparison with several variants of the Heston and Nandi model that shows the superiority of the Variance-Gamma innovations also from the point of view of historical MLE estimation.

# 1. Introduction

Several empirical studies have documented important departures from the assumption of normality of log-returns. Indeed skewness, kurtosis, serial correlation and time-varying volatilities are observed in financial time series. For this reason different models have been investigated in discrete and in continuous time.

---

[1] Dipartimento di Metodi Quantitativi, Università di Milano Bicocca, Italy. E-mail: lorenzo.mercuri@unimib.it

[2] Dipartimento di Metodi Quantitativi, Università di Milano Bicocca, Italy. E-mail: fabio.bellini@unimib.it

In continuous time, the Lévy processes seem to be a natural generalization of the Brownian motion. Indeed the Lévy process exhibits right-continuous sample paths with stationary and independent increments. Moreover, the marginal distribution can be easily identified by characteristic function (see for example Schoutens (2003) and the references therein). However, the Lévy processes usually represent an incomplete market and therefore we need to choose an equivalent martingale measure. The standard approach is based on Esscher transform or the Minimal Entropy Martingale Measure (see Hubalek and Sgarra (2006) for a survey and comparison of these measures).

Another way to capture the departure from normality is based on the concept of random time, introduced in finance by Clark (1973). A new process, namely the subordinated process, can be obtained from a primitive stochastic process by using an independent random time change process, referred to as a subordinator (usually an increasing Lévy process). The distribution of the resulting process is closely related to a mixture distribution. In particular, if we consider the time-changed Brownian motion, the distribution at time one is a Normal variance-mean mixture distribution. Some cases considered in the literature are the Variance-Gamma (see Madan and Seneta (1990)), the Normal Inverse Gaussian (see Barndorff-Nielsen (1995)) and the Hyperbolic and Generalized Hyperbolic distributions (see Barndorff-Nielsen (1977)).

As far as discrete time models are considered, the main classes are stochastic volatility models and Garch-like models.

In stochastic volatility models, the distribution of returns is specified indirectly by the structure of the model, indeed there exists a random variable V such that the conditional distribution of log-returns given V is known (usually normal). This kind of assumption is often made in continuous-time where the volatility also follows a diffusion process. The main drawback of this approach is that the stochastic volatility is an unobservable process and this gives rise to estimation difficulties.

Garch-like models explicitly model the conditional variance given the past observed returns and volatilities. For option pricing, the affine Garch models represent a suitable class, since they yield a closed form formula for option prices based on inverse Fourier transform (see Heston and Nandi (2000) for normal innovations, Christoffersen et al. (2006) for Inverse Gaussian innovations, Bellini and Mercuri (2007) for Gamma innovations and Mercuri (2008) for Tempered Stable innovations).

In this paper, we present a new discrete-time stochastic volatility model where the logreturns follow a conditional Variance-Gamma distribution. As we have already mentioned, the Variance-Gamma (VG henceforth) distribution belongs to the class of normal variance-mean mixtures and corresponds to a Gamma mixing density. In a static one period framework, the VG distribution has shown a good ability to reproduce stylized facts of the distribution of financial logreturns (see for example Madan and Seneta (1987)). Later these authors considered the VG process (Madan and Seneta (1990)), that is a Lévy process with VG increments and applied it to option pricing.

Our idea is to try to capture a time varying conditional distribution by means of a *time varying Gamma mixing density*. To this aim we will use for the distribution of the conditional variance a simple affine Garch model with Gamma innovations, in order to capture persistence of high levels of volatility and of volatility shocks. Moreover, in contrast with usual Garch models, we will be able to capture also the time dynamic of higher order moments in a relatively simple model.

In order to select a martingale measure we will use the conditional Esscher transform proposed by Buhlmann (1996) and widely applied in Garch-like models with non-normal innovation (see Siu et al. (2004)). The main advantage of this approach is that the conditional distribution of the log-returns is still a VG after the change of measure and the resulting dynamic for the mixing density is similar to the dynamic under the historical measure.

Another advantage of this model is that it allows a recursive procedure for the determination of the characteristic function of the logprice at maturity and hence a semianalytical option pricing based on inverse Fourier transform as in Heston (1993) and Carr-Madan (1999). Moreover this model encompasses the VG discrete time model and an affine Garch Gamma model proposed by Bellini and Mercuri (2007) as special cases.

In Section 2, we review some classical results of Normal variance-mean distribution and we focus on Variance Gamma distribution. In Section 3, we present our model and, following the approach proposed by Heston and Nandi (2000), we obtain a recursive procedure for characteristic function and we achieve the affine Garch model with Gamma innovations and the Variance Gamma model as special cases. In Section 4, we apply the conditional Esscher transform introduced and we obtain a closed form formula for option prices by inverse Fourier transform (see Heston 1993).

In Section 5, we show by means of a detailed comparison with the Heston and Nandi model the superiority of the Variance Gamma innovations also from the point of view of historical estimation. We acknowledge the important contribution of an anonymous referee that suggested the idea of approximating the density of the Variance Gamma innovations with a Gauss-Laguerre quadrature, that worked very well in practice.

In Section 6, the proposed model is calibrated on 738 daily closing prices of European options on S&P500 and compared with the Heston and Nandi (2000) model and with the Christoffersen, Heston and Jacobs (2006) model, with promising results.

## 2. The Variance-Gamma distribution

In this section we review the basic properties of the Variance-Gamma distribution, introduced in Finance by Madan and Seneta (1990). It belongs to the class of *normal variance-mean mixtures* introduced in Barndorff-Nielsen et al. (1982), that are defined as

$$Y := \mu_0 + \mu V + \sigma \sqrt{V} Z \qquad (1)$$

where $\mu_0$, $\mu \in R$, $\sigma \in [0, +\infty)$, $Z \sim N(0,1)$ and $V$ is a nonnegative random variable independent from $Z$.

Despite this parametrization is standard and convenient for applications, not all the parameters are identifiable; a simple way to overcome this problem is to impose $\sigma = 1$ or $E[V] = 1$.

In general, this distribution can be thought as a Brownian motion with drift $\mu$ and volatility $\sigma$, starting at $\mu_0$, and stopped at the random time $V$. In the $\mu = 0$ case, the bigger $V$ (in the usual stochastic order $\leq_{st}$), the more disperse $Y$ (in the convex order $\leq_{cx}$); see Theorem 3.A.3 in Shaked et al. (2006).

As remarked already in Barndorff-Nielsen et al. (1982), a possible theoretical motivation for the use of normal variance-mean mixtures comes from a generalized version of the central limit theorem with a random number of summands (Renyi 1960).

If the mixing distribution $V$ admits a density $g$, then the density of $Y$ is given by

$$f(y) = \frac{1}{\sqrt{2\pi\sigma^2}} \int_0^{+\infty} \frac{g(s)}{\sqrt{s}} \exp\left(-\frac{(y-\mu_0-\mu s)^2}{2\sigma^2 s}\right) ds \qquad (2)$$

and if $V$ admits a finite moment generating function $M_V$, then the m.g.f. of $Y$ is given by

$$M_Y(c) = \exp(c\mu_0) M_V\left(c\mu + \frac{c^2\sigma^2}{2}\right) \qquad (3)$$

Perhaps the most important example of normal variance-mean mixture is the Generalized Hyperbolic distribution (GH), that arise when $V$ has a Generalized Inverse Gaussian (GIG) distribution (see for example Prause (1999) and the references therein).

The *Variance-Gamma (VG) distribution* (Madan and Seneta (1990)) is the special case obtained by choosing a Gamma mixing density $V \sim \Gamma(a,b)$. The corresponding density function has not a simple analytic form (it admits however a representation in terms of Bessel functions of the second kind, see Madan et al. (1990) and the references therein), while the m.g.f. becomes

$$M_Y(c) = \exp(c\mu_0)\left(\frac{b}{b-c\mu-\frac{c^2\sigma^2}{2}}\right)^a \qquad (4)$$

showing that if $X_1,...,X_n$ are i.i.d. and VG distributed, then also their sum is of the same type with $\mu_0^{(n)} = n\mu_0$ and $a^{(n)} = na$, the other parameters remaining unchanged.

The first moments are given by

$$E[Y] = \mu_0 + \frac{a\mu}{b}$$

$$Var(Y) = \frac{a(\mu^2 + b\sigma^2)}{b^2}$$

$$skew(Y) = \frac{\mu(2\mu^2 + 3b\sigma^2)}{a^{\frac{1}{2}}(\mu^2 + b\sigma^2)^{\frac{3}{2}}} \qquad (5)$$

$$kurt(Y) = 3\left\{1 + \frac{2\mu^4 + b^2\sigma^4 + 4b\sigma^2\mu}{a(\mu^2 + b\sigma^2)^2}\right\}$$

where as already remarked for the general normal variance-mean mixture one of the 5 parameters $\mu_0$, $\mu$, $\sigma$, $a$, $b$ is redundant and will be fixed for definiteness. The remaining 4 parameters family is able to capture both skewness and kurtosis of the logreturns; from the definition (1) we see that if $\mu = 0$ then the distribution is symmetric around $\mu_0$, while from (5) we see that the asymmetry has the same sign as the parameter $\mu$.

The kurtosis is always greater than 3 and is a decreasing function of $a$. Moreover, in the symmetric, zero mean and $\sigma = 1$ case we get the remarkably simple formulas $Var[Y] = \frac{a}{b} = E[V]$ and $kurt(Y) = 3\left(1 + \frac{1}{a}\right)$.

The shape of the densities for different values of the parameters are reported in the following picture:

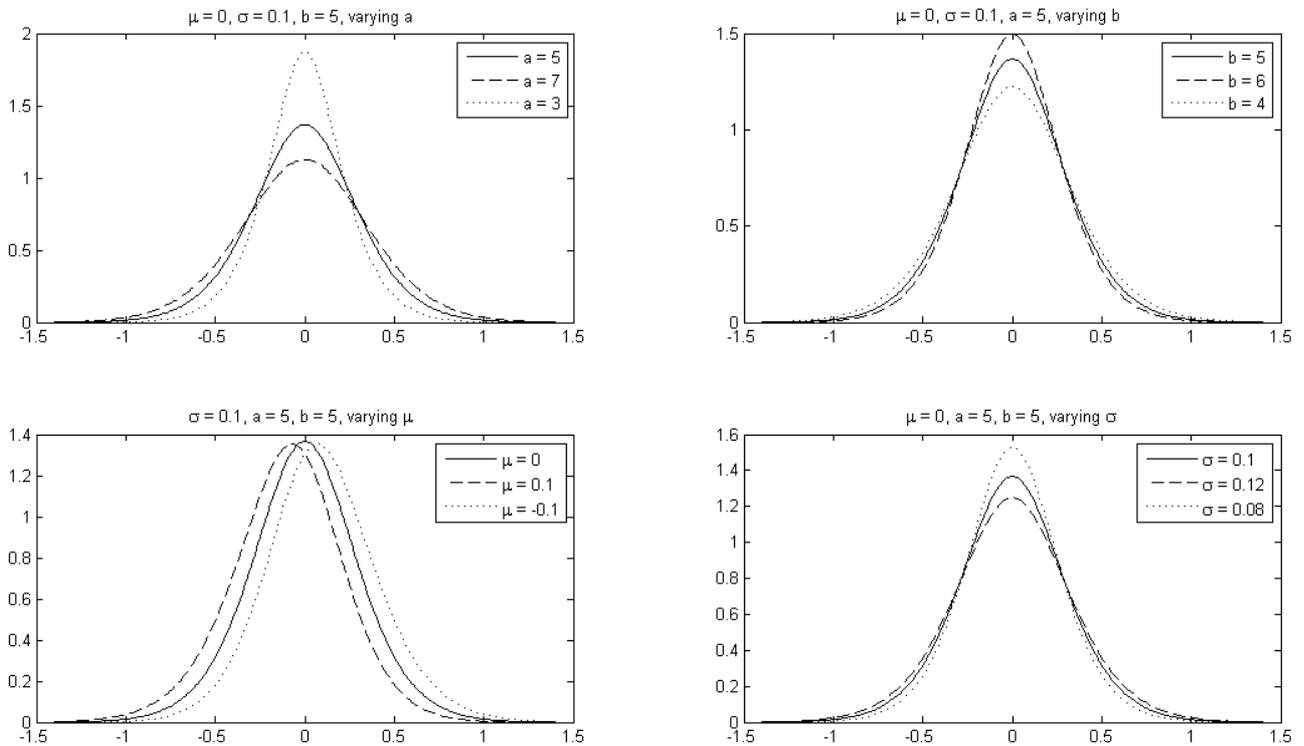

Fig. 1 VG densities for $\mu_0 = 0$ and different values of the parameters $a, b, \mu$ and $\sigma$.

Since if $V_1 \sim \Gamma(a_1, b)$ and $V_2 \sim \Gamma(a_2, b)$ with $a_1 \leq a_2$ implies $V_1 \leq_{st} V_2$ (see for example Muller et al. (2002)), from the preceding remark we see that the variance-gamma family is monotonic in the convex order $\leq_{cx}$ with respect to the parameter $a$, as is also evident in Fig.1 (upper left).

As shown by Madan and Seneta, the VG distribution seems to exhibit a good capability of fitting historical logreturns and has also becoming increasingly popular in option pricing; for example a VG pricing model is even implemented in the Bloomberg data providing systems (see for example Carr et al. (2007) and the references therein).

# 3. A dynamic VG model

In this section we propose a *conditional VG* model. The basic idea is to make the Gamma mixing distribution $V$ time dependent, by allowing its shape parameter to vary in a recursive fashion; hence we will assume that

$$\begin{cases} V_t \mid F_{t-1} \sim \Gamma(ah_t, 1) \\ h_t = \alpha_0 + \alpha_1 V_{t-1} + \beta_1 h_{t-1} \end{cases} \quad (6)$$

Thus we are going to build *a stochastic volatility model* in which the conditional volatility $V_t$ follows a Garch-like process with Gamma innovations. The recursive equation for $h_t$ is designed to capture some degree of persistence of high volatility periods, exactly in the same way as in the usual Garch(1,1) models. A shock in previous volatility will propagate to future volatility through the term $\alpha_1 V_{t-1}$; an high level of previous volatility will propagate to future volatility through the term $\beta_1 h_{t-1}$.

The complete model for the logreturns $Y_t$ then becomes

$$\begin{cases} Y_t = r + \lambda V_t + \sigma \sqrt{V_t} Z_t \\ V_t \mid F_{t-1} \sim \Gamma(ah_t, 1) \\ Z_t \sim N(0,1) \text{ i.i.d.} \\ h_t = \alpha_0 + \alpha_1 V_{t-1} + \beta_1 h_{t-1} \end{cases} \quad (7)$$

with $a, \alpha_0, \alpha_1, \beta_1 > 0$.

As we remarked at the end of the previous section, the bigger $h_t$, the bigger in the usual stochastic order the conditional distribution of $V_t$, the more dispersed in the convex order the distribution of $Y_t$.

We remark that in contrast to usual Garch models, and similarly to continuous time stochastic volatility models, there are here *two sources of risk*: the stochastic recurrence (6) for the volatility and the random shocks $Z_t$ for the logreturns. The idea is to try to model variance dynamics by itself, without a coupling term proportional to squared past logreturns, but with a persistency term proportional to past variance.

The model then results in a conditional VG distribution for the logreturns, with a time varying mixing density $V_t$. The conditional variance is then given by

$$Var_{t-1}(Y_t) = (\sigma^2 + \lambda^2) a h_t \tag{8}$$

and since the parameter $a$ is redundant we see that choosing

$$a = \frac{1}{\sigma^2 + \lambda^2} \tag{9}$$

we have simply $Var_{t-1}(Y_t) = h_t$, exactly as in Garch models. But in contrast to Garch models, the dynamic features of this model are not limited to a time dependent conditional variance, but also involve time dependent higher order moments.

With this parametrization, the model depends on the six parameters $r$, $\lambda$, $\sigma$, $\alpha_0$, $\alpha_1$, $\beta_1$. In the next section we will see that the martingale condition will imply a relationship between $\lambda$ and $\sigma$, reducing the number of the parameters to 5. Clearly, when $\alpha_0 = \alpha_1 = 0$ and $\beta_1 = 1$ we recover a model with i.i.d. variance-gamma logreturns, while in the case $\sigma = 0$ we recover an affine Garch gamma model that has been studied in Bellini and Mercuri (2007).

We now compute recursively the m.g.f. of $\log(S_T)$ following the same approach as Heston and Nandi (2000). Recalling that

$$S_T = S_t \exp\left(\sum_{k=t}^{T} Y_k\right)$$

and defining the conditional m.g.f. of the terminal stock price $\log(S_T)$ as

$$\varphi_t(c) = E_t[S_T^c] \tag{10}$$

where $E_t[\cdot]$ is the time-t conditional expectation, we claim that $\varphi_t(c)$ has the form

$$\varphi_t(c) = S_t^c \exp(A(t;T,c) + B(t;T,c) h_{t+1}) \tag{11}$$

where the quantities $A(t;T,c)$ and $B(t;T,c)$ will be computed recursively. Assuming that equation (11) holds at time $t+1$ by the iteration law of conditional expectations we get

$$\varphi_t(c) = E_t[E_{t+1}[\varphi_{t+1}(c)]] =$$
$$= E_t[\exp(c\log(S_{t+1}) + A(t+1;T,c) + B(t+1;T,c)h_{t+2})] =$$
$$= S_t^c \exp(cr + A(t+1;T,c) + \alpha_0 B(t+1;T,c) + \beta_1 B(t+1;T,c)h_{t+1}) *$$
$$* E_t[\exp(c\lambda + \alpha_1 B(t+1;T,c))V_{t+1} + c\sigma\sqrt{V_{t+1}}Z_{t+1}]$$

substituting the expression of the m.g.f. of a variance-gamma we get:

$$\varphi_t(c) = S_t^c \exp[cr + A(t+1;T,c) + \alpha_0 B(t+1;T,c) + \beta_1 B(t+1;T,c)h_{t+1}] *$$
$$* \left[1 - \left(c\lambda + \alpha_1 B(t+1;T,c) + \frac{c^2\sigma^2}{2}\right)\right]^{-ah_{t+1}}$$

where $A(t;T,c)$ and $B(t;T,c)$ follow the recursions

$$\begin{cases} A(t;T,c) = cr + A(t+1;T,c) + \alpha_0 B(t+1;T,c) \\ B(t;T,c) = \beta_1 B(t+1;T,c) - a\log\left[1 - \left(c\lambda + \alpha_1 B(t+1;T,c) + \frac{c^2\sigma^2}{2}\right)\right] \end{cases} \quad (12)$$

with terminal conditions

$$\begin{cases} A(T;T,c) = 0 \\ B(T;T,c) = 0 \end{cases} \quad (13)$$

We can check that in the i.i.d. case ($\alpha_0 = \alpha_1 = 0$, $\beta_1 = 1$) the explicit solution is

$$\begin{cases} A(t;T,c) = c(T-t)r \\ B(t;T,c) = -a(T-t)\log\left[1 - c\lambda - \frac{c^2\sigma^2}{2}\right] \end{cases} \quad (14)$$

and the conditional m.g.f. is given by

$$\varphi_t(c) = S_t^c \exp(c(T-t)r)\left(1 - c\lambda - \frac{c^2\sigma^2}{2}\right)^{-ah(T-t)} \quad (15)$$

where $h$ is the common value of $h_1,...,h_T$; as expected we get for the logreturn a variance-gamma distribution where the mixing gamma has a shape factor $ah(T-t)$.

We remark that similar models can be constructed by using alternative characterizations for the dynamics of the mixing distribution, based on the Inverse Gaussian Garch model (see Christoffersen et al. (2006) or Tempered Stable Garch model (see Mercuri (2008)), while essentially retaining the same degree of analytical tractability.

## 4. Option pricing

The proposed model is not in general a martingale; in order to use it for option pricing purposes it is necessary to find an equivalent martingale measure. As it is customary with discrete time models with continuous innovations, the model is incomplete. The standard way of constructing an equivalent martingale measure is by means of the conditional Esscher transform proposed by Buhlmann et al. (1996) and applied in Garch framework in several papers (see for example Siu et al. (2004) for a case with Gamma innovations).

The conditional m.g.f. of log-returns can be written as:

$$M_t(c) = E_{t-1}[\exp(cY_t)] = \exp(cr)\left(1 - c\lambda - \frac{c^2\sigma^2}{2}\right)^{-ah_t}$$

The conditional Esscher change of measure is a product of "local" change of measures $\Lambda_t$ (see Buhlmann et al. (1996)) or Shiryaev et al. (1999).

$$\frac{dQ}{dP} = \prod_{t=1}^{T} \Lambda_t, \text{ with } \Lambda_t \geq 0 \text{ a.s. and } E_{t-1}(\Lambda_t) = 1 \qquad (16)$$

where $\Lambda_t$ is exponential in the logreturns

$$\Lambda_t = \frac{\exp(\theta_t^* Y_t)}{E_{t-1}[\exp(\theta_t^* Y_k)]} \qquad (17)$$

and where the Esscher parameter $\theta_t^*$ is obtained by solving the conditional Esscher equation

$$\frac{M_t(\theta_t^* + 1)}{M_t(\theta_t^*)} = e^r \qquad (18)$$

that gives

$$\theta^* = -\left(\frac{\lambda}{\sigma^2} + \frac{1}{2}\right) \quad (19)$$

that does not depend on $t$.

The conditional moment generating function of $Y_t$ under $Q$ is then given by

$$M_t^Q(c) = \frac{M_t(c+\theta^*)}{M_t(\theta^*)} = \exp(cr)\left(1 - c\frac{\sigma^4 4}{(\sigma^4 - 4\lambda^2 - 8\sigma^2)} + c^2\frac{\sigma^4 4}{(\sigma^4 - 4\lambda^2 - 8\sigma^2)}\right)^{-ah_t} \quad (20)$$

that can be written as

$$\exp(cr)\left(1 - c\lambda_Q - \frac{c^2\sigma_Q^2}{2}\right)^{-ah_t}$$

by posing

$$\lambda_Q = \frac{4\sigma^4}{(\sigma^4 - 4\lambda^2 - 8\sigma^2)} \quad \text{and} \quad \frac{(\sigma_Q)^2}{2} = -\lambda_Q \quad (21)$$

from which we see that the risk neutral distribution of the innovations is again a Variance Gamma.

Under $Q$ the log-returns dynamics is then given by

$$\begin{cases} Y_t = r - \frac{\sigma_Q^2}{2}V_t + \sigma_Q\sqrt{V_t}Z_t \\ V_t \mid F_{t-1} \sim \Gamma(ah_t, 1) \\ Z_t \sim N(0,1) \text{ i.i.d.} \\ h_t = \alpha_0 + \alpha_1 V_{t-1} + \beta_1 h_{t-1} \end{cases} \quad (22)$$

Introducing the risk neutral conditional variance

$$h_t^Q := Var_{t-1}^Q(Y_t) = a[\sigma_Q^2 + \lambda_Q^2]h_t \quad (23)$$

and letting

$$\begin{cases} \alpha_0^Q = a[\sigma_Q^2 + \lambda_Q^2]\alpha_0 & \alpha_1^Q = a[\sigma_Q^2 + \lambda_Q^2]\alpha_1 \\ \beta_1^Q = \beta_1 & a^Q = \frac{1}{\sigma_Q^2 + \lambda_Q^2} \end{cases} \quad (24)$$

the model becomes:

$$\begin{cases} Y_t^Q = r - \dfrac{\sigma_Q^2}{2} V_t^Q + \sigma^Q \sqrt{V_t^Q} Z_t \\ V_t^Q \mid F_{t-1} \sim \Gamma(a^Q h_t^Q, 1) \\ Z_t \sim N(0,1) \\ h_t^Q = \alpha_0^Q + \alpha_1^Q V_{t-1} + \beta_1^Q h_{t-1}^Q \end{cases} \quad (25)$$

and it is identified by four parameters $\sigma_Q$, $\alpha_0^Q$, $\alpha_1^Q$, $\beta_1^Q > 0$.

In the following figure we compare the density of the log-price under the real measure with the corresponding density under the Esscher martingale measure for the values of the parameters $\lambda = 0$, $\sigma = 0.1$, $\lambda_Q = -0.005$, $\sigma_Q = 0.1001$, $a = 3$, $\alpha_0 = 0.05$, $\alpha_1 = 0.12$, $\beta_1 = 0.08$ and $h_0 = 0.15$.

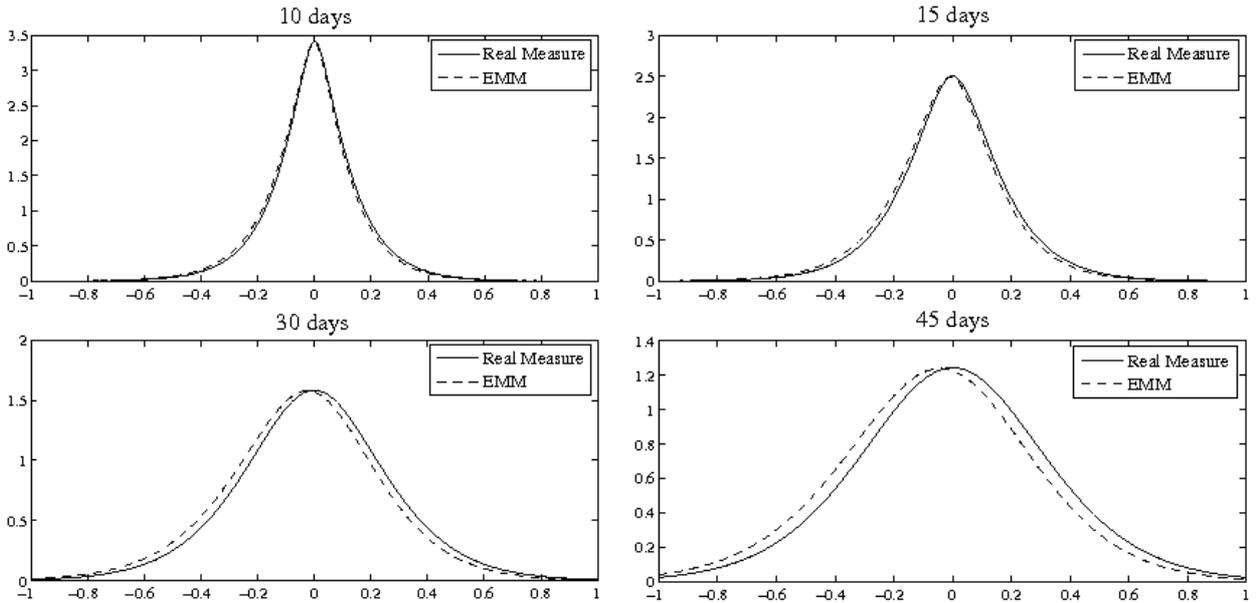

Fig. 2 Comparison between real and risk neutral densities for different time horizons.
Both distributions are obtained by means of inverse Fourier transform.

In order to check the correctness of our recursive semianalytical procedure for option pricing, we compare the resulting prices with Montecarlo prices obtained by means of $N = 100000$ simulations of the risk neutral model, for different maturities and for different levels of moneyness. The parameters chosen for the simulations are $\lambda_Q = -0.005$, $\sigma_Q = 0.1001$, $a = 3$, $\alpha_0 = 0.05$, $\alpha_1 = 0.12$, $\beta_1 = 0.08$ and $h_0 = 0.15$.

We see that the prices obtained by means of inverse Fourier Transform (FT) are typically close to the Montecarlo price and in all the considered cases belongs to the 95% confidence interval computed according to the methodology of Boyle (1977). The results are reported in the following table:

| T | Strike | MC | LB (95%) | UB (95%) | FT |
|---|--------|------|----------|----------|------|
|     | 0.9  | 0.1626 | 0.1611 | 0.1640 | 0.1632 |
|     | 0.95 | 0.1344 | 0.1330 | 0.1357 | 0.1350 |
| 1 m | 1    | 0.1102 | 0.1089 | 0.1114 | 0.1108 |
|     | 1.05 | 0.0898 | 0.0886 | 0.0909 | 0.0905 |
|     | 1.1  | 0.0728 | 0.0717 | 0.0739 | 0.0736 |
|     | 0.9  | 0.2041 | 0.2019 | 0.2062 | 0.2048 |
|     | 0.95 | 0.1790 | 0.1770 | 0.1810 | 0.1797 |
| 2 m | 1    | 0.1567 | 0.1547 | 0.1586 | 0.1573 |
|     | 1.05 | 0.1369 | 0.1351 | 0.1388 | 0.1375 |
|     | 1.1  | 0.1195 | 0.1178 | 0.1213 | 0.1201 |
|     | 0.9  | 0.2363 | 0.2337 | 0.2390 | 0.2373 |
|     | 0.95 | 0.2130 | 0.2105 | 0.2156 | 0.2139 |
| 3 m | 1    | 0.1919 | 0.1895 | 0.1944 | 0.1926 |
|     | 1.05 | 0.1729 | 0.1705 | 0.1752 | 0.1734 |
|     | 1.1  | 0.1557 | 0.1534 | 0.1580 | 0.1561 |

Tab. 1 Comparison between Montecarlo and semianalytical prices in the DVG model

# 5. Historical estimation and comparison with the Heston and Nandi model

In the preceding sections we saw that our model leads to an efficient semianalytic procedure for option pricing based on the recursive computation of the characteristic function of the underlying, as for example in the Heston and Nandi model. The aim of this section is to assess the relevance of VG innovations from an historical point of view, in comparison with the HN model that is based on normal innovations.

The main problem in historical estimation is that the VG innovations don't have a simple analytic density; we are seriously indebted with an anonymous referee that gave us the key suggestion of approximating the VG density by means of a Gauss-Laguerre quadrature of the integral

$$f(y) = \int_0^{+\infty} \frac{g(s)}{\sqrt{2\pi\sigma^2 s}} \exp\left(-\frac{(y-\mu_0-\mu s)^2}{2\sigma^2 s}\right) ds \qquad (26)$$

where $g(s)$ is a gamma density. This numerical procedure actually worked very well and allowed us to obtain quite stable historical estimates; from a more statistical point of view we could say that the VG innovations are approximated by a finite mixture of normals.

We present a comparative study of the historical estimation of 5 increasingly complicated models: Heston and Nandi without asymmetry in the dynamic of the variance, Heston and Nandi, VG without asymmetry in the dynamic of the variance and with symmetric innovations, VG with symmetric innovations and asymmetric variance dynamics, VG. With our notations, the HN model can be written as:

$$\begin{cases} Y_t = r + \lambda V_t + \sqrt{V_t} Z_t \\ V_t = \alpha_0 + \alpha_1(Z_{t-1} - \gamma)^2 + \beta_1 V_{t-1} \\ Z_t \text{ i.i.d.} \end{cases} \qquad (27)$$

and the 5 considered models are: model (HN) with $\gamma = 0$ and normal innovations (MOD1), model (HN) with normal innovations (MOD2), model (HN) with $\gamma = 0$ and standardized symmetric VG innovations (MOD3), model (HN) with standardized symmetric VG innovations (MOD4), model (HN) with standardized VG innovations (MOD5).

By comparing the estimation results we are able to evaluate separately the relevance of the asymmetry in the variance dynamic (due to the parameter $\gamma$) and the asymmetry in the distribution of the innovations (due to their standardized VG distribution in place of the standard normal).

As it was suggested by an anonymous referee, the simplest way to parametrize the standardized VG innovations is to choose $\mu_0 = -\mu := -\alpha$, $a = b := k$, $\sigma = \sqrt{1 - \frac{\alpha^2}{k}}$; we call it a $SVG(\alpha, k)$ distribution.

Moreover it is easy to check that if $Y \sim SVG(\alpha, k)$, then

$$Skew(Y) = \frac{\alpha(3k - \alpha^2)}{k^2}, \quad kurt(Y) = 3\left\{1 + \frac{2\alpha^4 + (k - \alpha^2)(k + 4\alpha - \alpha^2)}{k^3}\right\}$$

The likelihood is built by means of a Gauss-Laguerre quadrature of the integral

$$f(y) = \int_0^{+\infty} \frac{k^k s^{k-1} e^{-ks}}{\Gamma(s)} \frac{1}{\sqrt{2\pi\sigma^2 s}} \exp\left(-\frac{(y - \mu_0 - \mu s)^2}{2\sigma^2 s}\right) ds =$$

$$= \int_0^{+\infty} \frac{u^{k-1} e^{-u}}{\Gamma(u/k)} \frac{1}{\sqrt{2\pi\sigma^2 \frac{u}{k}}} \exp\left(-\frac{\left(y - \mu_0 - \mu \frac{u}{k}\right)^2}{2\sigma^2 \frac{u}{k}}\right) du \qquad (28)$$

$$= \int_0^{+\infty} e^{-u} h(u) du \cong \sum_{i=i}^{n} w_i h(x_i)$$

where the abscissas $x_i$ are the roots of the Laguerre polynomials $L_n(x)$ and the weights $w_i$ can be easily calculated as

$$w_i = \frac{x_i}{(n+1)^2 L_{n+1}^2(x_i)}$$

(see for example Abramowitz and Stegun (1972)); as suggested by the referee we chose $n = 10$.

The dataset is composed by approximately 1100 logreturns of the SP500 index, ranging from 3/01/2006 to 18/05/2010; we deliberately included the period of exceptional market conditions that arose across the financial crisis. The SPX logreturms have been corrected by the dividend yield computed by Bloomberg and the risk free rate has been extracted from Bloomberg's C079 curve. The estimation results are presented in the following tables:

|            | MOD1                    | MOD2                    |
|------------|-------------------------|-------------------------|
| $\lambda$  | 1,20 (1,72)             | 2,09 (1,86)             |
| $\alpha_0$ | 1,31E-05 (3,00E-06)     | 2,20E-16 (5,51E-13)     |
| $\alpha_1$ | 2,39E-05 (4,62E-06)     | 1,22E-05 (1,97E-06)     |
| $\beta_1$  | 0,81 (0,03)             | 0,89 (0,02)             |
| $\gamma$   | -                       | 57,74 (4,35E-02)        |
| logL       | 3211,01                 | 3296,50                 |

Tab. 2 Estimated parameters for the 2 considered models with normal innovations

|            | MOD3                    | MOD4                    | MOD5                    |
|------------|-------------------------|-------------------------|-------------------------|
| $\lambda$  | 5,84 (1,76)             | 5,19 (1,23E-03)         | 2,59 (7,17E-04)         |
| $\alpha_0$ | 2,0E-16 (4,6E-12)       | 2,22E-16 (7,29E-15)     | 2,22E-16 (1,01E-14)     |
| $\alpha_1$ | 1,09E-05 (3,71E-06)     | 4,49E-06 (4,79E-07)     | 4,81E-06 (5,06E-07)     |
| $\beta_1$  | 0,94 (0,02)             | 0,62 (0,04)             | 0,63 (0,04)             |
| $\gamma$   | -                       | 277,95 (9,68E-05)       | 264,56 (1,26E-04)       |
| $k$        | 1,24 (0,17)             | 1,31 (1,86E-01)         | 1,30 (1,84E-01)         |
| $\alpha$   | -                       | -                       | -2,19E-01 (4,89E-02)    |
| logL       | 3341,53                 | 3379,00                 | 3389,40                 |

Tab. 3 Estimated parameters for the 3 considered

The first table refers to normal innovations, while the second to SVG innovations. For all models we find a positive risk premium on the underlying, and the estimation error greatly reduces when moving from normal to VG innovations. The parameters $\alpha_0$ and $\alpha_1$ are typically quite close to $0$, and in some cases their 95% asymptotic confidence interval actually includes $0$ as a possible value. The parameter $\beta$ is always significantly different from $0$, as is the parameter $\gamma$, modelling the asymmetry in the variance dynamic. The parameter $k \cong 1,3$ of the symmetric SVG innovations of MOD3 and MOD4 corresponds to $kurt(Y) \cong 5,1$ thus indicating conditional tails heavier than the Gaussian - but not extremely heavy. In the general SVG case we obtain $skew(Y) \cong -0,5$, $kurt(Y) \cong 3,65$.

In order to assess the relative improvement in the maximum loglikelihood values when moving from MOD1 to MOD5, we perform a standard likelihood ratio test; the results are reported in the following table and shows that in all cases the improvement in considering the more complex model (that is with VG innovations and asymmetry in the variance dynamic) is substantial.

|  | LR statistic | $p$ value |
|---|---|---|
| MOD1 / MOD2 | 170,98 | 4.51E - 39 |
| MOD1 / MOD3 | 261,05 | 1.01E - 58 |
| MOD2 / MOD4 | 165,00 | 9.14E - 38 |
| MOD4 / MOD5 | 20,80 | 5.09E - 06 |

Tab. 4 Likelihood ratio test of the 5 considered models

## 6. Calibrations and comparisons

The aim of this section is to investigate the ability of the dynamic VG model to reproduce market option prices.

We compare its calibration performance with the Heston and Nandi (2000) model (HN), and with the Christoffersen, Heston and Jacobs (2006) model (CHJ) that is based on Inverse Gaussian innovations. Moreover, in order to highlight the importance of a time-varying mixing distribution, we compare the fitting of the DVG model with two submodels obtained by letting $\alpha_0 = \alpha_1 = 0$, $\beta_1 = 1$ (constant parameters VG model) and by letting $\sigma = 0$ (that corresponds to an affine Garch Gamma model).

At the moment we follow a purely calibration approach (as for example in Bakshi et al. (1997)). The alternative could be to implement a "mixed" historical/calibration procedure as in Heston and Nandi (2000) or as in Christoffersen et al. (2006), but this could lead to a more difficult assessment of the calibration results; we leave it for further research. A recent discussion of the relative merits of "mixed" historical/calibration methods versus purely historical MLE and purely calibration NLS procedures can be found in Menn and Rachev (2009).

We consider European options on the S&P500 index quoted at the CBOE. The dataset is composed by 738 daily closing prices from 12/23/08 to 02/17/09. On each day we have on the average approximately 20 closing prices of options with the three different maturities February, March and April. Moneyness ranges from 0.975 and 1.025 .

For each of the 5 considered models (DVG, HN, CHJ, VG, GG) we calibrate everyday the corresponding parameters (that are 3 in the VG case, 4 in the DVG, HN and GG case, 5 in the CHJ case). For the VG case we adopt the parametrization that can be found for example in Schoutens (2003).

In order to perform the calibrations we adopt two different criteria: the (dollar) root mean squared error

$$\$RMSE = \sqrt{\frac{1}{n}\sum_{i=1}^{n}\left(C_i^{theo}(\hat{\theta}) - C_i^{mkt}\right)^2}$$

and the percentage root mean squared error, that is defined as

$$\%RMSE = \sqrt{\frac{1}{n}\sum_{i=1}^{n}\left(\frac{C_i^{theo}(\hat{\theta}) - C_i^{mkt}}{C_i^{mkt}}\right)^2}$$

where $\hat{\theta}$ is the vector of the parameters.

It is well known that the choice of the loss function might affect the estimated parameters and the quality of the calibrations; for the explorative character of our analysis we stick to the two most common loss functions, other choices being squared errors on implied volatilities, or weighted mean squared errors.

It is clear that the $\$RMSE$ criterion gives implicitly more weight to deep in the money options, that have high prices, while the $\%RMSE$ criterion gives more weight to very low price, deep out of the money options. For a detailed discussion of these issues see for example Christoffersen et al. (2004).

The minimizations are performed by the Newton-Raphson algorithm implemented in the **fmincon** routine in Matlab environment.

The $\$RMSE$ and the $\%RMSE$ of the calibrated models are reported in the following two figures:

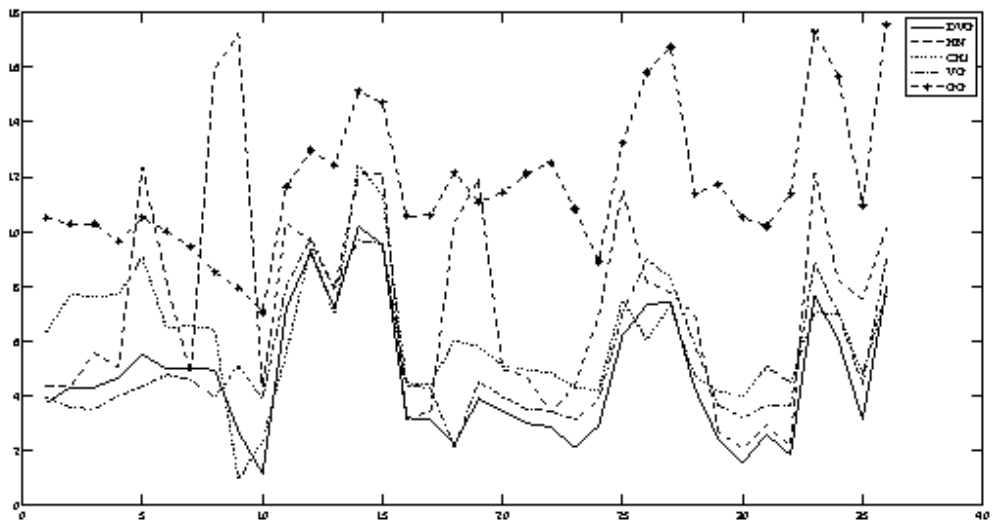

Fig. 3 Comparison of $\$RMSE$ of the 5 considered model

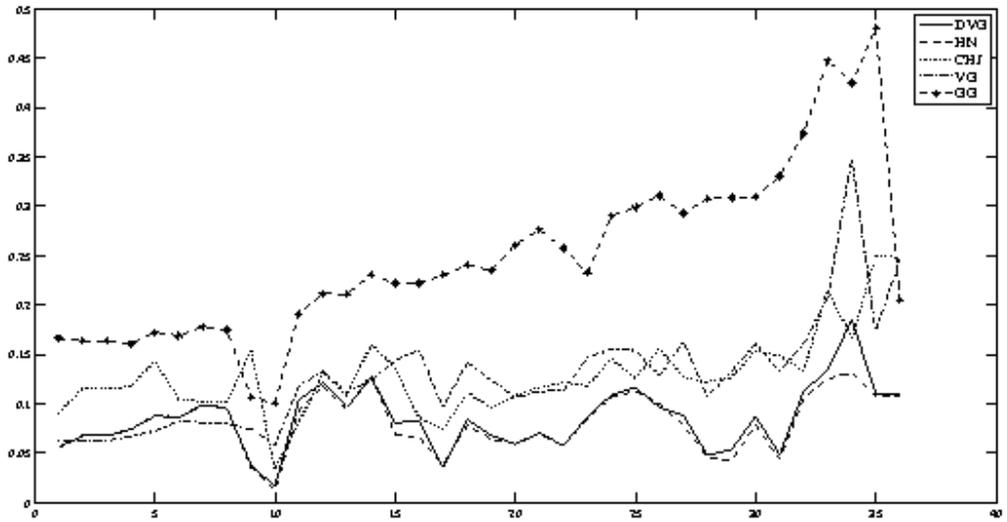

Fig. 4 Comparison of $\%RMSE$ of the 5 considered model

We see that for what concerns the $\$RMSE$ criterion the DVG model seems to perform typically better in the considered period, while for what concerns the $\%RMSE$ we have a substantial equality of the performance of DVG and HN models. In the following table we report the averages of daily $\$RMSE$ and daily $\%RMSE$.

|  | DVG | HN | CHJ | VG | GG |
|---|---|---|---|---|---|
| $\$RMSE$ | 4.81 | 7.50 | 6.18 | 5.48 | 11.76 |
| $\%RMSE$ | 9.88% | 9.55% | 13.93% | 13.35% | 25.08% |

Tab. 5 Average of daily $\$RMSE$ and daily $\%RMSE$

The three parameters VG model does not perform badly, but it is systematically beaten by the DVG model for what concerns $\%RMSE$. The CHJ model seems to perform quite worse, but some of its problems have already been pointed out in Christoffersen et al. (2006).

In order to explore the stability of the fitted parameters, we report the graph of the DVG and HN parameters fitted with $\$RMSE$:

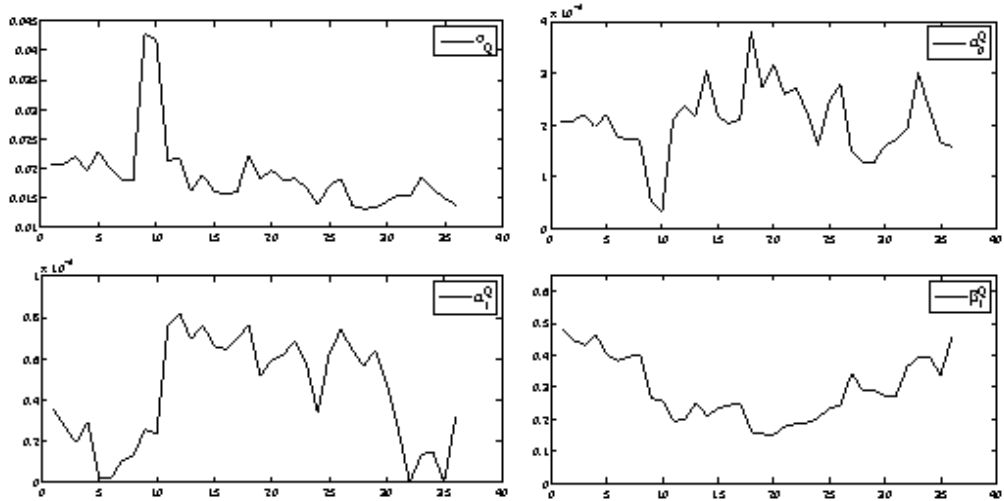

Fig. 5 Estimated parameters for the DVG model

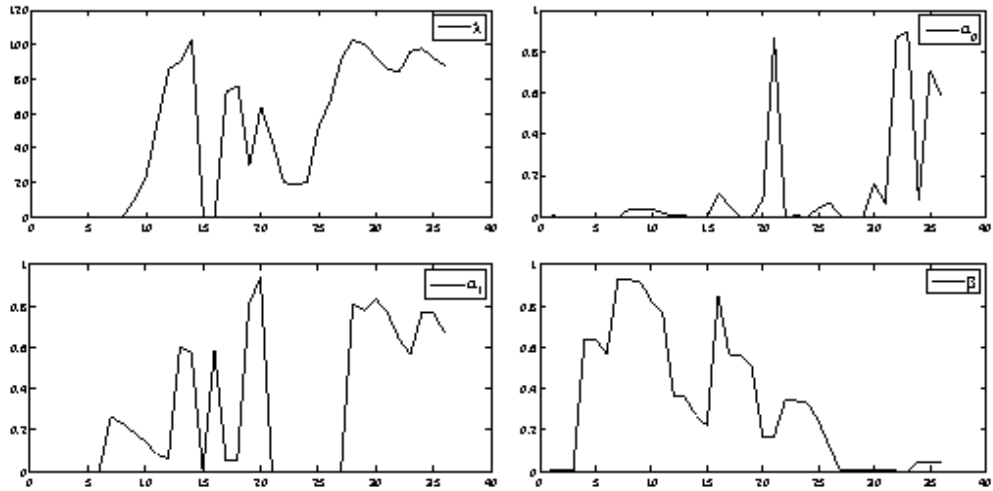

Fig. 6 Estimated parameters for the HN model

from where we see that the DVG model seems to lead to more stable estimations.

These empirical analysis show that the DVG model is quite promising and represents an improvement over the HN model, while retaining the same degree of analytical tractability.

The results of the historical estimation in the preceding section also show the superiority of the VG innovations with respect to the normal innovations of the HN model.

# Acknowledgements

We would like to thank two anonymous referees for their helpful comments and suggestions. All remaining errors are responsibility of the authors.

# References

Abramowitz, M., Stegun I. A. (1972). *Handbook of Mathematical Functions With Formulas, Graphs, and Mathematical Tables* Dover Publications.

Bakshi, G., Cao C., Chen Z. (1997) "Empirical performance of alternative option pricing models" *Journal of Finance*, 52, 2003 - 2049